\documentstyle[12pt,epsfig]{article}
\newcommand{\bea}{\begin{eqnarray}}
\newcommand{\eea}{\end{eqnarray}}
\newcommand{\be}{\begin{equation}}
\newcommand{\ee}{\end{equation}}

\newcommand{\nn}{\nonumber}

\begin{document}
\begin{center}
{\large\bf Model-independent approach as a tool of studying chiral symmetry}
\bigskip

{\bf Yu.S.~Surovtsev}$^{a,}$\footnote{E-mail address: surovcev@thsun1.jinr.ru},
{\bf D.~Krupa}$^b$, {\bf M.~Nagy}$^b$
\bigskip

$^a${\it Bogoliubov Laboratory of Theoretical Physics, Joint Institute for
Nuclear Research, Dubna 141 980, Moscow Region, Russia},\\
$^b${\it Institute of Physics, Slov.Acad.Sci., D\'ubravsk\'a cesta 9,
842 28 Bratislava, Slovakia}
\end{center}

\begin{abstract}
A simultaneous analysis (in the model-independent approach) of the $\pi\pi$
scattering (from the threshold to 1.9 GeV) and of the
$\pi\pi\to K\overline{K}$ process (from the threshold to $\sim$ 1.46 GeV)
is carried out in channel with $I^GJ^{PC}=0^+0^{++}$. The existence of the
$f_0(665)$ state with properies of the $\sigma$-meson is proved. An
indication for the glueball nature of the $f_0(1500)$ is obtained. A minimum
scenario of the simultaneous description of the processes
$\pi\pi\to\pi\pi,K\overline{K}$ does without the ${f_0}(1370)$ resonance.
A version including this state is also considered. In this case the
${f_0}(1370)$ resonance has the dominant $s{\bar s}$ component (the ratio of
its coupling constant with the $\pi\pi$ channel to the one with the
$K\overline{K}$ channel is 0.12). The coupling constants of the observed
states with the $\pi\pi$ and $K\overline{K}$ systems and the $\pi\pi$ and
$K\overline{K}$ scattering lengths are obtained.
A parameterless description of the $\pi\pi$ background is given by allowance
for the left-hand branch-point in the uniformizing variable. On the basis of
this and of the fact that all the adjusted parameters for the $\pi\pi$
scattering are only the positions of poles describing resonances, it is
concluded that our model-independent approach is a valuable tool for studying
the realization schemes of chiral symmetry. The existence of the $f_0(665)$
state and the obtained $\pi\pi$-scattering length ($a_0^0\approx 0.27
m_{\pi^+}^{-1}$) seem to suggest the linear realization of chiral symmetry.
\end{abstract}

\section{Introduction}
The discovered states in the scalar sector and their properties do not allow
one up to now to make up the scalar $q{\bar q}$ nonet. Difficulties in
understanding these mesons are related to the theoretical explanation of
their properties and obtaining a model-independent information on these
objects as wide multichannel states. Earlier, we have shown \cite{KMS-nc96}
that an inadequate description of multichannel states gives not only their
distorted parameters when analyzing data but also can cause the fictitious
states when one neglects important channels. In this report, we shall show
that the large background, which earlier one has obtained in various analyses
of the $s$-wave $\pi\pi$ scattering \cite{PDG-00}, hides, in reality, the
$\sigma$-meson and the influence of the left-hand branch-point. The state
with the $\sigma$-meson properties is required by most models (like the
linear $\sigma$-models and the Nambu -- Jona-Lasinio models \cite{NJL}) for
spontaneous breaking of chiral symmetry. We obtain also an indication for the
glueball nature of the $f_0(1500)$ and for the dominant $s{\bar s}$ component
of the $f_0(1370)$ state.

A model-independent information on multichannel states and their QCD nature
can be obtained only on the basis of the first principles (analyticity and
unitarity) immediately applied to analyzing experimental data and using the
mathematical fact that a local behaviour of analytic functions, determined on
the Riemann surface, is governed by the nearest singularities on all sheets.
Earlier, we have proposed that method for 2- and 3-channel resonances and
developed the concept of standard clusters (poles on the Riemann surface) as
a qualitative characteristic of a state and a sufficient condition of its
existence \cite{KMS-nc96,KMS-cz88}. We outline this below for the 2-channel
case. Then we analyze simultaneously experimental data on the processes
$\pi\pi\to\pi\pi,K\overline{K}$ in the channel with $I^GJ^{PC}=0^+0^{++}$.
On the basis of obtained pole clusters for resonances, their coupling
constants with the considered channels and the scattering lengths, compared
with the results of various models, we conclude on the nature of the observed
states and on the mechanism of chiral-symmetry breaking.

\section{Two-Coupled-Channel Formalism}
We consider the coupled processes of $\pi\pi$ and $K\overline{K}$ scattering
and $\pi\pi\to K\overline{K}$. Therefore, we have the two-channel $S$-matrix
determined on the 4-sheeted Riemann surface. The $S$-matrix elements
$S_{\alpha\beta}$, where $\alpha,\beta=1(\pi\pi), 2(K\overline{K})$, have the
right-hand (unitary) cuts along the real axis of the $s$-variable complex
plane ($s$ is the invariant total energy squared), starting at $4m_\pi^2$ and
$4m_K^2$, and the left-hand cuts, beginning at $s=0$ for $S_{11}$ and at
$4(m_K^2-m_\pi^2)$ for $S_{22}$ and $S_{12}$. The Riemann-surface sheets are
numbered according to the signs of analytic continuations of the channel
momenta ~$k_1=(s/4-m_\pi^2)^{1/2}$, ~$k_2=(s/4-m_K^2)^{1/2}$~ as follows:
signs $({\mbox{Im}}k_1,{\mbox{Im}}k_2)=++,-+,--,+-$ correspond to the
sheets I, II, III, IV.

To obtain the resonance representation on the Riemann surface, we express
analytic continuations of the matrix elements to the unphysical sheets
$S_{\alpha\beta}^L$ ($L=II,III,IV$) in terms of those on the physical sheet
$S_{\alpha\beta}^I$, which have only zeros (beyond the real axis),
corresponding to resonances. Using the reality property of the analytic
functions and the 2-channel unitarity, one can obtain
\begin{eqnarray} \label{S_L}
&&S_{11}^{II}=\frac{1}{S_{11}^I},\qquad ~~~~S_{11}^{III}=
\frac{S_{22}^I}{\det S^I}, \qquad
S_{11}^{IV}=\frac{\det S^I}{S_{22}^I},\nonumber\\
&&S_{22}^{II}=\frac{\det S^I}{S_{11}^I},\qquad S_{22}^{III}=\frac{S_{11}^I}
{\det S^I},\qquad S_{22}^{IV}=\frac{1}{S_{22}^I},\\
&&S_{12}^{II}=\frac{iS_{12}^I}{S_{11}^I},\qquad ~~~S_{12}^{III}=
\frac{-S_{12}^I}{\det S^I},\qquad S_{12}^{IV}=
\frac{iS_{12}^I}{S_{22}^I},\nonumber
\end{eqnarray}
Here  $\det S^I=S_{11}^I S_{22}^I-(S_{12}^I)^2;
(S_{12}^I)^2=-s^{-1}\sqrt{(s-4m_\pi^2)(s-4m_K^2)}F(s)$; in the limited energy
interval, $F(s)$ is proportional to the squared product of the coupling
constants of the considered state with channels 1 and 2. Formulae (\ref{S_L})
immediately give the resonance representation by poles and zeros on the
4-sheeted Riemann surface. Here one must discriminate between three types of
resonances -- which are described ({\bf a}) by a pair of complex conjugate
poles on sheet II and, therefore, by a pair of complex conjugate zeros on the
Ist sheet in $S_{11}$; ({\bf b}) by a pair of conjugate poles on sheet IV
and, therefore, by a pair of complex conjugate zeroes on sheet I in $S_{22}$;
({\bf c}) by one pair of conjugate poles on each of sheets II and IV, that is
by one pair of conjugate zeroes on the physical sheet in each of matrix
elements $S_{11}$ and $S_{22}$.

As seen from (\ref{S_L}), to the resonances of types ({\bf a}) and ({\bf b})
one has to make correspond a pair of complex conjugate poles on sheet III
which are shifted relative to a pair of poles on sheet II and IV ,
respectively. For the states of type ({\bf c}), there are two pairs of
conjugate poles on sheet III. Thus, we arrive at the notion of three standard
pole-clusters which represent two-channel bound states of quarks and gluons.
The cluster must be a compact formation (on the $s$-plane) of size typical
for strong interactions ($<150-200$ MeV). The cluster type must be defined by
a state nature. We can distinguish a bound state of particles (colourless
objects) and bound states of quarks and gluons. The resonance, coupled
relatively more strongly with the 1st ($\pi\pi$) channel than with the 2nd
($K\overline{K}$) one, is described by the pole cluster of type ({\bf a});
in the opposite case, by the cluster of type ({\bf b}) ({\it e.g.}, if it has
a dominant $s\overline{s}$ component); finally, a flavour singlet
({\it e.g.}, glueball),{\it i.e.}, the state having approximately equal
coupling constants with the considered members of the nonet is represented by
the cluster of type ({\bf c}) as a necessary condition.

For the simultaneous analysis of experimental data on the coupled processes
it is convenient to use the Le Couteur-Newton relations \cite{LN} expressing
the $S$-matrix elements of all coupled processes in terms of the Jost matrix
determinant $d(k_1,k_2)$, the real analytic function with the only
square-root branch-points at $k_i=0$ \cite{Kato}. To take account of these
right-hand branch-points, the corresponding uniformizing variable should be
used. Earlier, this was done by us in the 2-channel consideration
\cite{KMS-cz88} with the uniformizing variable
~$z=(k_1+k_2)/(m_K^2-m_\pi^2)$,
which was proposed in Ref.\cite{Kato} and maps the 4-sheeted Riemann surface
with two unitary cuts, starting at the points $4m_\pi^2$ and $4m_K^2$, onto
the plane. (Note that other authors have also applied the parametrizations
with using the Jost functions at analyzing the $s$-wave $\pi\pi$ scattering
in the one-channel approach \cite{Bohacik} and in the two-channel one
\cite{MP-92}. In latter work, the uniformizing variable $k_2$ has been used,
therefore, their approach cannot be employed near the $\pi\pi$ threshold.)

When analyzing the processes $\pi\pi\to \pi\pi,K\overline{K}$ by the above
methods in the 2-channel approach, two resonances ($f_0 (975)$ and
$f_0 (1500)$) are found to be sufficient for a satisfactory description
($\chi^2/\mbox{ndf}\approx1.00$). However, in this case, a large
$\pi\pi$-background has been obtained. A character of the representation of
the background (the pole of second order on the imaginary axis on sheet II
and the corresponding zero on sheet I) suggests that a wide light state is
possibly hidden in the background. To check this, one must work out the
background in some detail.

Now we take into account also the left-hand branch-point at $s=0$ in the
uniformizing variable
\begin{equation} \label{v}
v=\frac{m_K\sqrt{s-4m_\pi^2}+m_\pi\sqrt{s-4m_K^2}}{\sqrt{s(m_K^2-m_\pi^2)}}.
\end{equation}
This variable maps the 4-sheeted Riemann surface, having (in addition to two
above-indicated unitary cuts) also the left-hand cut starting at the point
$s=0$, onto the $v$-plane. (Note that the analogous uniformizing variable
has been used, {\it e.g.}, in Ref. \cite{Meshch} at studying the forward
elastic $p{\bar p}$ scattering amplitude.)
\begin{figure}
\centering
\epsfig{file=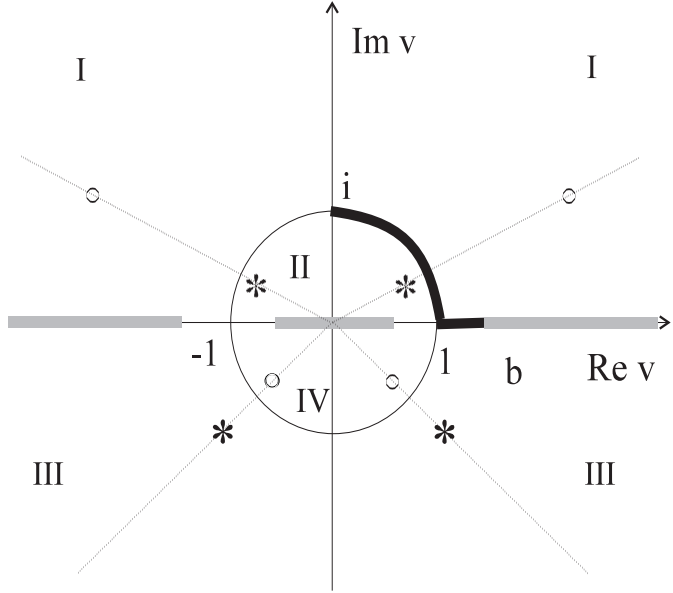, width=9cm}
\caption{
Uniformization plane for the \protect$\pi\pi$-scattering amplitude.
}
\label{fig:v.plane}
\end{figure}
In Fig.\ref{fig:v.plane}, the plane of the uniformizing variable $v$ for the
$\pi\pi$-scattering amplitude is depicted. The Roman numerals(I,\ldots, IV)
denote the images of the corresponding sheets of the Riemann surface; the
thick line represents the physical region; the points i, 1 and
$b=\sqrt{(m_K+m_{\pi})/(m_K-m_{\pi})}$ correspond to the $\pi\pi,
K\overline{K}$ thresholds and $s=\infty$, respectively; the shaded intervals
$(-\infty,-b],~[-b^{-1},b^{-1}],~[b,\infty)$ are the images of the
corresponding edges of the left-hand cut. The depicted positions of poles
($*$) and of zeros ($\circ$) give the representation of the type ({\bf a})
resonance in $S_{11}$.

On $v$-plane the Le Couteur-Newton relations are \cite{KMS-cz88,Kato}
\begin{equation} \label{v:C-Newton}
S_{11}=\frac{d(-v^{-1})}{d(v)},\quad S_{22}=\frac{d(v^{-1})}{d(v)},
\quad S_{11}S_{22}-S_{12}^2=\frac{d(-v)}{d(v)}.
\end{equation}
Then, the condition of the real analyticity implies $ ~d(-v^*)=d^* (v)~$
for all $v$, and the unitarity needs the following relations to hold true for
the physical $v$-values: $ ~|d(-v^{-1})|\leq |d(v)|$, $~~ |d(v^{-1})|\leq
|d(v)|,~$  $~|d(-v)|=|d(v)|$.

The $d$-function that on the $v$-plane already does not possess branch-points
is taken as $~d=d_B d_{res}$,
where ~$d_B=B_{\pi}B_K$; $B_{\pi}$ contains the possible remaining
$\pi\pi$-background contribution, related to exchanges in crossing channels;
$B_K$ is that part of the $K\overline{K}$ background which does not
contribute to the $\pi\pi$-scattering amplitude. The most considerable part
of the background of the considered coupled processes related to the
influence of the left-hand branch-point at $s=0$ is taken already in the
uniformizing variable $v$ (\ref{v}) into account. The function $d_{res}(v)$
represents the contribution of resonances, described by one of three types of
the pole-zero clusters, {\it i.e.}, except for the point $v=0$, it consists
of the zero s of clusters:
\begin{equation} \label{d_res}
d_{res}= v^{-M}\prod_{n=1}^{M} (1-v_n^* v)(1+v_n v),
\end{equation}
where $M$ is the number of pairs of the conjugate zeros.

\section{Analysis of experimental data}
We analyze simultaneously the available experimental data on the
$\pi\pi$-scattering \cite{Hyams} and the process $\pi\pi\to K\overline{K}$
\cite{Wickl} in the channel with $I^GJ^{PC}=0^+0^{++}$. As data, we use the
results of phase analyses which are given for phase shifts of the amplitudes
($\delta_1$ and $\delta_{12}$) and for moduli of the $S$-matrix elements
$\eta_1$ (the elasticity parameter) and $\xi$:
\begin{equation} \label{del.eta}
S_a =\eta_a e^{2i\delta_a}~~~~(a=1,2),\qquad S_{12} =i\xi e^{i\delta_{12}}.
\end{equation}
("1" denotes the $\pi\pi$ channel, "2" -- $K\overline{K}$).
The 2-channel unitarity condition gives
~~$\eta_1=\eta_2=\eta,~~ \xi=(1-\eta^2)^{1/2},~~ \delta_{12} =
\delta_1+\delta_2$.\\
We have taken the data on the $\pi\pi$ scattering from the threshold up to
1.89 GeV. Then, comparing experimental data for $\xi$ with calculated values
using experimental points for $\eta$, one can see that the 2-channel
unitarity takes place to about 1.4 GeV.

\begin{figure}
\centering
\epsfig{file=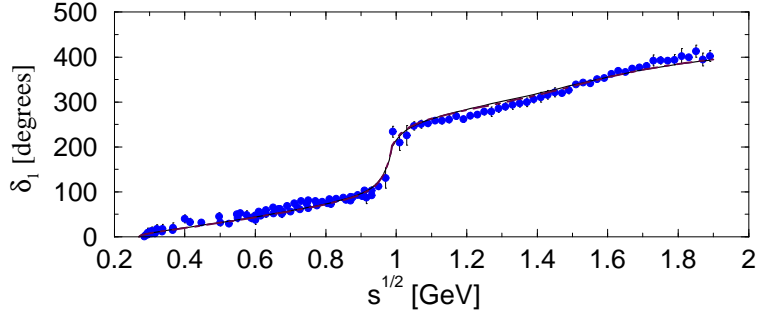, width=10cm}
\caption{
The energy dependence of the phase shift \protect$\delta_1$ of the $\pi\pi$
scattering amplitude obtained on the basis of a simultaneous analysis of the
experimental data on the coupled processes $\pi\pi\to\pi\pi,K\overline{K}$
in the channel with $I^GJ^{PC}=0^+0^{++}$ for both versions -- without
and with the $f_0(1370)$ state: the solid curve corresponds to version 1;
the dot-dashed one, version 2 (for \protect$\delta_1$ the both curves
practically coincide). The data on the $\pi\pi$ scattering are taken from
Refs. \cite{Hyams}.
}
\label{fig:pipi.phas}
\end{figure}

\begin{figure}
\centering
\epsfig{file=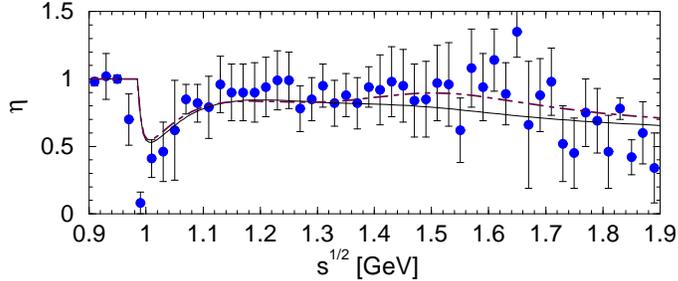, width=9cm}
\caption{
The same as in Fig. \ref{fig:pipi.phas} but for the elasticity parameter
$\eta$.
}
\label{fig:pipi.mod}
\end{figure}

To obtain the satisfactory description of the $s$-wave $\pi\pi$ scattering
from the threshold to 1.89 GeV, we have taken $B_\pi=1$, and three
multichannel resonances turned out to be sufficient: the two ones of the type
({\bf a}) ($f_0 (665)$ and $f_0 (980)$) and $f_0 (1500)$ of the type
({\bf c}). Therefore, in eq.(\ref{d_res}) $M=8$ and the following zero
positions on the $v$-plane, corresponding to these resonances have been
established in this situation with the parameterless description of the
background:
\begin{eqnarray}
{\rm for} ~~f_0 (665):
~~&&v_1=1.36964+0.208632i,\qquad ~~v_2 =0.921962-0.25348i,\nonumber\\
{\rm for} ~~f_0 (980):~
&&v_3=1.04834+0.0478652i,\qquad ~v_4 =0.858452-0.0925771i,\nonumber\\
{\rm for} ~~f_0 (1500):
&&v_5=1.2587+0.0398893i,\qquad ~~~v_6 =1.2323-0.0323298i,\nonumber\\
&&v_7=0.809818-0.019354i,\qquad ~v_8 =0.793914-0.0266319i.\nonumber
\end{eqnarray}
For $\delta_1$ and $\eta$, 113 and 50 experimental points \cite{Hyams},
respectively, are used; when rejecting the points at energies 0.61, 0.65, and
0.73 GeV for $\delta_1$ and at 0.99, 1.65, and 1.85 GeV for $\eta$, which
give an anomalously large contribution to $\chi^2$, we obtain for
$\chi^2/\mbox{ndf}$ the values 2.7 and 0.72, respectively; the total
$\chi^2/\mbox{ndf}$ in the case of the $\pi\pi$ scattering is 1.96.
The corresponding curves (solid) demonstrating the quality of these fits are
shown in Figs. \ref{fig:pipi.phas} and \ref{fig:pipi.mod}.
\begin{figure}
\centering
\epsfig{file=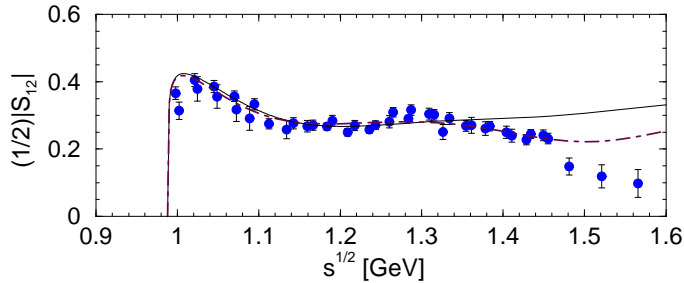, width=9cm}
\caption{
The energy dependences of the \protect($|S_{12}|$) for both soluitons: the
solid curve corresponds to solution (1); the dot-dashed one, solution (2).
The data on the process \protect$\pi\pi\to K\overline{K}$ are taken from Refs.
\cite{Wickl}.
}
\label{fig:piK.mod}
\end{figure}
With the presented picture, the satisfactory description for the modulus
($\xi$) of the $\pi\pi\to K\overline{K}$ matrix element is given from the
threshold to $\sim$ 1.4 GeV (Fig.\ref{fig:piK.mod}, the solid curve).
Here 35 experimental points \cite{Wickl} are used;
$\chi^2/\mbox{ndf}\approx 1.11$ when eliminating the points at energies
1.002, 1.265, and 1.287 GeV (with especially large contribution to $\chi^2$).
However, for the phase shift $\delta_{12}(s)$, slightly excessive curve is
obtained. Therefore, keeping the {\it parameterless} description of the
$\pi\pi$ background, one must take into account the part of the
$K\overline{K}$ background that does not contribute to the
$\pi\pi$-scattering amplitude. Note that the variable $v$ is uniformizing for
the $\pi\pi$-scattering amplitude, {\it i.e.}, on the $v$-plane, $S_{11}$ has
no cuts, however, the amplitudes of the $K\overline{K}$ scattering and
$\pi\pi\to K\overline{K}$ process do have the cuts on the $v$-plane, which
arise from the left-hand cut on the $s$-plane, starting at the point
$s=4(m_K^2-m_\pi^2)$. This left-hand cut will be neglected in the
Riemann-surface structure, and the contribution on the cut will be taken into
account in the $K\overline{K}$ background as a pole on the real $s$-axis on
the physical sheet in the sub-$K\overline{K}$-threshold region. On the
$v$-plane, this pole gives two poles on the unit circle in the upper
half-plane, symmetric to each other with respect to the imaginary axis, and
two zeros, symmetric to the poles with respect to the real axis, {\it i.e.},
at describing the process $\pi\pi\to K\overline{K}$, one additional parameter
is introduced, say, a position $p$ of the zero on the unit circle. Therefore,
for $B_K$ we take the form
\begin{equation} \label{B_K}
B_K=v^{-4}(1-pv)^4(1+p^*v)^4.
\end{equation}
The fourth power in (\ref{B_K}) is stipulated by the following. First, a pole
on the real $s$-axis on the physical sheet in $S_{22}$ is accompanied by a
pole in sheet II at at the same $s$-value (as it is seen from eqs.
(\ref{S_L})); on the $v$-plane, this implies the pole of second order. Second,
for the $s$-channel process $\pi\pi\to K\overline{K}$, the crossing $u$- and
$t$-channels are the $\pi-K$ and $\overline{\pi}-K$ scattering (exchanges in
these channels give contributions on the left-hand cut). This results in the
additional doubling of the multiplicity of the indicated pole on the
$v$-plane. The expression (\ref{B_K}) does not contribute to $S_{11}$,
{\it i.e.}, the parameterless description of the $\pi\pi$ background is kept.
A satisfactory description of the phase shift $\delta_{12}(\sqrt{s})$
(Fig.\ref{fig:piK.phas}) is obtained to $\sim$ 1.52 GeV with the value of
$p=0.948201+0.31767i$ (this corresponds to the pole position on the
$s$-plane at $s=0.434 {\rm GeV}^2$).
\begin{figure}
\centering
\epsfig{file=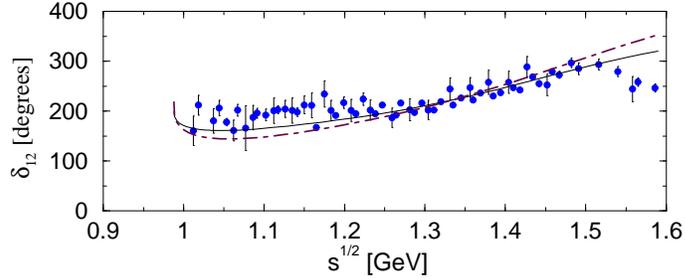, width=9cm}
\caption{
The same as in Fig.~\ref{fig:piK.mod} but for the phase shift
\protect($\delta_{12}$).
}
\label{fig:piK.phas}
\end{figure}
Here 59 experimental points \cite{Wickl} are considered; $\chi^2/\mbox{ndf}
\approx 3.05$ when eliminating the points at energies 1.117, 1.247, and 1.27
GeV (with especially large contribution to $\chi^2$). The total
$\chi^2/\mbox{ndf}$ for four analyzed quantities to describe the processes
$\pi\pi\to\pi\pi,K\overline{K}$ is 2.12; the number of adjusted parameters is
17.

In Table~\ref{tab:clusters}, the obtained poles on the corresponding sheets
of the Riemann surface are shown on the complex energy plane ($\sqrt{s_r}=
={\rm E}_r-i\Gamma_r$). Since, for wide resonances, values of masses and
widths are very model-dependent, it is reasonable to report characteristics
of pole clusters which must be rather stable for various models.
\begin{table}
\centering
\caption{
Pole clusters for obtained resonances in solution (1).
}
\vskip0.3truecm
\begin{tabular}{|c|rl|rl|rl|rl|rl|rl|}
\hline
{} & \multicolumn{2}{c|}{$f_0 (665)$} & \multicolumn{2}{c|}{$f_0(980)$}
& \multicolumn{2}{c|}{$f_0(1500)$} \\
\cline{2-7}
Sheet & \multicolumn{1}{c}{E, MeV} & \multicolumn{1}{c|}{$\Gamma$, MeV}
& \multicolumn{1}{c}{E, MeV} & \multicolumn{1}{c|}{$\Gamma$, MeV}
& \multicolumn{1}{c}{E, MeV} & \multicolumn{1}{c|}{$\Gamma$, MeV}\\
\hline
II & 610$\pm$14 & 620$\pm$26 & 988$\pm$5 & 27$\pm$8 & 1530$\pm$25
& 390$\pm$30 \\
\hline
III & 720$\pm$15 & 55$\pm$9 & 984$\pm$16 & 210$\pm$22 & 1430$\pm$35
& 200$\pm$30 \\
{} & {} & {} & {} & {} & 1510$\pm$22~ & 400$\pm$34 \\
\hline
IV & {} & {} & {} & {} & 1410$\pm$24 & 210$\pm$38  \\
\hline \end{tabular}
\label{tab:clusters}
\end{table}

Now we calculate the coupling constants of the obtained states with the
$\pi\pi$ and $K\overline{K}$ systems through the residues of amplitudes at
the pole on sheet II, expressing the $T$-matrix via the $S$-matrix as
$~S_{ii}=1+2i\rho_i T_{ii}$, $~ S_{12}=2i\sqrt{\rho_1\rho_2} T_{12}$,
where $\rho_i=\sqrt{(s-4m_i^2)/s}$, and taking the resonance part of the
amplitude in the form $~T_{ij}^{res}=\sum_r g_{ir}g_{rj}D_r^{-1}(s),$
where $D_r(s)$ is an inverse propagator ($D_r(s)\propto s-s_r$).
The obtained values of the coupling constants are given in
Table~\ref{tab:constants} (we denote the coupling constants with the
$\pi\pi$ and $K\overline{K}$ systems through $g_1$ and $g_2$, respectively).
\begin{table}
\centering
\caption{
Coupling constants of the observed states with the $\pi\pi$ ($g_1$)
and $K\overline{K}$ ($g_2$) systems in solution (1).
}
\vskip0.3truecm
\begin{tabular}{|l|l|l|l|} \hline
{}  & $f_0(665)$ & $f_0(980)$ & $f_0(1500)$\\ \hline
$g_{1}$, GeV & ~$0.7477\pm 0.095$~ & ~$0.1615 \pm 0.03$~ & ~$0.899 \pm 0.09
3$~\\
\hline
$g_2$, GeV & ~$0.834\pm 0.1$~ & ~$0.438 \pm 0.028$~ & {}\\
\hline
\end{tabular}
\label{tab:constants}
\end{table}

In this 2-channel approach, there is no point in calculating the coupling
constant of the $f_0(1500)$ state  with the $K\overline{K}$ system, because
the 2-channel unitarity is valid only to 1.4 GeV, and, above this energy,
there is a considerable disagreement between the calculated amplitude
modulus $S_{12}$ and experimental data.

We see that a minimum scenario of the simultaneous description of the
processes $\pi\pi\to\pi\pi,K\overline{K}$ does not require the ${f_0}(1370)$
resonance. Therefore, if this meson exists, it must be relatively more weakly
coupled to the $\pi\pi$ channel than to the $K\overline{K}$ one, {\it i.e.},
it should be described by the pole cluster of type ({\bf b}) (this would
testify to the dominant $s{\bar s}$ component in this state). To confirm
quantitatively this qualitative conclusion \cite{skn-app,skn-prd} that is
distinct from the one of other works \cite{PDG-00}, we consider the 2nd
solution including the ${f_0}(1370)$ of type ({\bf b}) in addition to three
above-observed resonances. In the figures, this solution is represented by
the dot-dashed curves. The description of the $\pi\pi$ scattering from the
threshold up to 1.89 GeV is practically the same as without the $f_0(1370)$:
$\chi^2/\mbox{ndf}$ for two quantities $\delta_1$ and $\eta$ is 2.01.
The description of experimental data is improved a little for $|S_{12}|$
which is described now up to $\sim$ 1.46 GeV. For this quantity, we consider
now 41 experimental points \cite{Wickl}; $\chi^2/\mbox{ndf}\approx 0.92$.
However, on the whole, the description is even worse as compared with the
1st solution: the total $\chi^2/\mbox{ndf}\approx 2.93$ for four analyzed
quantities describing the processes $\pi\pi\to\pi\pi,K\overline{K}$ (cf.
2.12 for the 1st case). The number of adjusted parameters is 21, where they
all are positions of the poles describing resonances except a single one
related to the $K\overline{K}$ background which is $p=0.976745+0.214405i$
(this corresponds to the pole on the $s$-plane at $s=0.622 {\rm GeV}^2$).
Let us indicate the obtained zero positions on the $v$-plane, corresponding
to considered resonances in version 2:
\bea
{\rm for} ~~f_0 (665):
~~&&v_1=1.36783+0.212659i,\qquad ~~v_2 =0.921962-0.25348i,\nn\\
{\rm for} ~~f_0 (980):~
&&v_3=1.04462+0.0479703i,\qquad ~v_4 =0.858452-0.0925771i,\nn\\
{\rm for} ~~f_0 (1370):~
&&v_5=1.22783-0.0483842i,\qquad ~v_6 =0.802595-0.0379537i,\nn\\
{\rm for} ~~f_0 (1500):
&&v_7=1.2587+0.0398893i,\qquad ~~~v_8 =1.24837-0.0358916i,\nn\\
&&v_9=0.804333-0.0179899i,\qquad ~v_10 =0.795579-0.0253985i.\nn
\eea

In Table~\ref{tab:clusters2}, the obtained poles on the corresponding sheets
of the Riemann surface are shown on the complex energy plane ($\sqrt{s_r}=
{\rm E}_r-i\Gamma_r$).

\begin{table}
\centering
\caption{
Pole clusters for obtained resonances in solution (2).
}
\vskip0.3truecm
\begin{tabular}{|c|rl|rl|rl|rl|rl|rl|rl|rl|}
\hline
{} & \multicolumn{2}{c|}{$f_0(665)$}& \multicolumn{2}{c|}{$f_0(980)$}
& \multicolumn{2}{c|}{$f_0(1370)$} & \multicolumn{2}{c|}{$f_0(1500)$} \\
\cline{2-9}
Sheet
& \multicolumn{1}{c}{E, MeV} & \multicolumn{1}{c|}{$\Gamma$, MeV}
& \multicolumn{1}{c}{E, MeV} & \multicolumn{1}{c|}{$\Gamma$, MeV}
& \multicolumn{1}{c}{E, MeV} & \multicolumn{1}{c|}{$\Gamma$, MeV}
& \multicolumn{1}{c}{E, MeV} & \multicolumn{1}{c|}{$\Gamma$, MeV}\\
\hline
II & 610$\pm$14 & 610$\pm$26 & 986$\pm$5 & 25$\pm$8 & {} & {} & 1530$\pm$22
& 390$\pm$28 \\
\hline
III & 720$\pm$15 & 55$\pm$9 & 984$\pm$16 & 210$\pm$25 & 1340$\pm$21
& 380$\pm$25 & 1490$\pm$30 & 220$\pm$25 \\
{} & {} & {} & {} & {} & {} & {} & 1510$\pm$22~ & 370$\pm$30 \\
\hline
IV & {} & {} & {} & {} & 1330$\pm$18 & 270$\pm$20 & 1490$\pm$20 & 300$\pm$35\\
\hline \end{tabular}
\label{tab:clusters2}
\end{table}

When calculating the coupling constants, we must take, for the ${f_0}(1370)$
state, the residues of amplitudes at the pole on sheet IV. In
Table~\ref{tab:constants2}, the obtained values of coupling constants are
shown. We see that the ${f_0}(980)$ and especially the ${f_0}(1370)$ are
coupled essentially more strongly to the $K\overline{K}$ system than
to the $\pi\pi$ one. This tells about the dominant $s{\bar s}$ component
in the ${f_0}(980)$ state and especially in the ${f_0}(1370)$ one.
\begin{table}
\centering
\caption{
Coupling constants of the observed states with the $\pi\pi$ ($g_1$) and
$K\overline{K}$ ($g_2$) systems in solution (2).
}
\vskip0.3truecm
\begin{tabular}{|l|l|l|l|l|} \hline
{}  & $f_0(665)$ & $f_0(980)$ & $f_0(1370)$ & $f_0(1500)$\\ \hline
$g_{1}$, GeV & ~$0.724\pm 0.095$~ & ~$0.153 \pm 0.03$~ & ~$0.11\pm 0.03~$
& ~$0.866 \pm 0.09$~\\
\hline
$g_2$, GeV & ~$0.582\pm 0.1$~ & ~$0.521\pm 0.028$~ & ~$0.91\pm 0.04$~ & {}\\
\hline
\end{tabular}
\label{tab:constants2}
\end{table}

Let us indicate also scattering lengths calculated for both solutions. For
the $K\overline{K}$ scattering, we obtain
$$ a_0^0(K\overline{K})=-1.188\pm 0.13+(0.648\pm 0.09)i,~[m_{\pi^+}^{-1}];
~~~({\rm  solution (1)}),$$
$$a_0^0(K\overline{K})=-1.497\pm 0.12+(0.639\pm 0.08)i,~[m_{\pi^+}^{-1}];
~~~({\rm solution (2)}).$$
The presence of the imaginary part in $a_0^0(K\overline{K})$ reflects the
fact that, already at the threshold of the $K\overline{K}$ scattering, other
channels ($2\pi,4\pi$ etc.) are opened. We see that the real part of
$K\overline{K}$ scattering length is very sensitive to the existence of
the ${f_0}(1370)$ state.

In Table~\ref{tab:pipi.length}, we compare our results for the $\pi\pi$
scattering length $a_0^0$ obtained for both solutions with results of
other works both theoretical and experimental ones.
\begin{table}
\centering
\caption{
Comparison of results of various works for the $\pi\pi$ scattering
length $a_0^0$.
}
\vskip0.3truecm
\begin{tabular}{|c|l|l|} \hline
$a_0^0, ~m_{\pi^+}^{-1}$ & ~~~~~~~References & ~~~~~~~~~~~~~~~~~Remarks \\
\hline
$0.27\pm 0.06$ (1)& our paper & model-independent approach \\
$0.266$~~~~~~~~(2)&{}&{}\\
\hline
$0.26\pm 0.05$ & L. Rosselet et al.\cite{Hyams} & analysis of the decay
$K\to\pi\pi e\nu$ \\
{} & {} & using Roy's model\\
\hline
$0.24\pm 0.09$ & A.A. Bel'kov et al.\cite{Hyams} & analysis of
$\pi^-p\to\pi^+\pi^-n$ \\
{} & {} & using the effective range formula\\
\hline
$0.23$ & S. Ishida et al.\cite{Ishida} & modified analysis of $\pi\pi$
scattering \\
{} & {} & using Breit-Wigner forms \\
\hline
$0.16$ & S. Weinberg \cite{Weinberg} & current algebra (nonlinear
$\sigma$-model) \\
\hline
$0.20$ & J. Gasser, H. Leutwyler \cite{Gasser} & one-loop corrections,
nonlinear\\
{} & {} & realization of chiral symmetry \\
\hline
$0.217$ & J. Bijnens at al.\cite{Bijnens} & two-loop corrections, nonlinear\\
{} & {} & realization of chiral symmetry  \\
\hline
$0.26$ & M.K. Volkov \cite{Volkov} & linear realization of chiral symmetry
\\
\hline
$0.28$ & A.N. Ivanov, N.I. Troitskaya \cite{Ivanov} & a variant of chiral
theory with\\
{} & {} & linear realization of chiral symmetry
\\
\hline
\end{tabular}
\label{tab:pipi.length}
\end{table}

We see that our results correspond to the linear realization of chiral
symmetry.

We have here presented model-independent results: the pole positions,
coupling constants, and scattering lengths. Masses and widths of these
states that should be calculated from the obtained pole positions and
coupling constants are highly dependent on the used model. Let us demonstrate
this.

If the state $f_0(665)$ is the $\sigma$-meson, then from the known relation
$$g_{\sigma\pi\pi}=\frac{m_\sigma^2-m_\pi^2}{\sqrt{2}f_{\pi^0}}$$
(here $f_{\pi^0}$ is the constant of the weak decay of the $\pi^0$:
$f_{\pi^0}=93.1$ MeV) we obtain ~$m_\sigma\approx 342$ MeV. That small
value of the $\sigma$-mass can be explained in part by the mixing with the
$f_0(980)$ state \cite{Volk-Yud}. If we take the resonance part of the
amplitude as
$$T^{res}=\frac{\sqrt{s}\Gamma}{m_\sigma^2-s-i\sqrt{s}\Gamma,}$$
we obtain $m_\sigma\approx 850$ MeV and $\Gamma\approx 1240$ MeV.

\section{Conclusions}
In the model-independent approach consisting in the
immediate application to the analysis of experimental data of first
principles (analyticity-causality and unitarity), a simultaneous description
of the isoscalar $s$-wave channel of the processes $\pi\pi\to \pi\pi,
K\overline{K}$ from the thresholds to the energy values, where the 2-channel
unitarity is valid, is obtained with three states ($f_0(665), f_0 (980)$ and
$f_0(1500)$) being sufficient.
A parameterless description of the $\pi\pi$ background is given by allowance
for the left-hand branch-point in the proper uniformizing variable.
It is shown that the large $\pi\pi$-background, usually obtained in various
analyses, combines in reality the influence of the left-hand branch-point and
the contribution of a wide resonance at $\sim$ 665 MeV. Thus, a
model-independent confirmation of the state denoted in the PDG issues by
$f_0(400-1200)$ \cite{PDG-00} is obtained.
This is the $\sigma$-meson required by majority of models for spontaneous
breaking of chiral symmetry. Note also that a light $\sigma$-meson is needed,
for example, for an explanation of $K\to\pi\pi$ transitions using a Dyson --
Schwinger model \cite{M.Ivan}.
We emphasize that we have given, in fact, the first real proof of the
$\sigma$-meson existence, because our analysis is based only on the first
principles ({\it i.e.},it is dynamical-free) and on the mathematical fact
that a local behaviour of analytic functions determined on the Riemann
surface is governed by the nearest singularities on all sheets and makes no
additional assumptions about the the $\pi\pi$ background. For these reasons,
the obtained results are rather model-independent.
A multichannel state is represented by one of the standard clusters (poles on
the Riemann surface) which is a qualitative characteristic of a state and a
sufficient condition of its existence. The pole cluster gives the main effect
of a multichannel state. The cluster type must be defined by a state nature.
The compactness of a cluster and smoothness of the background are criteria of
the description being correct.

On the basis of the fact that all the adjusted parameters of describing the
$\pi\pi$ scattering are positions of poles corresponding to resonances
(because the parameterless description of the $\pi\pi$ background is
achieved), we conclude that our model-independent approach is a valuable tool
for studying the realization schemes of chiral symmetry.
The existence of $f_0(665)$ and the obtained $\pi\pi$-scattering length
($a_0^0(\pi\pi)\approx 0.27~m_{\pi^+}^{-1}$) suggest the linear realization
of chiral symmetry.

The discovery of the $f_0(665)$ state solves one important mystery of the
scalar-meson family that is related to the Higgs boson of the hadronic
sector. This is a result of principle, because the schemes of the nonlinear
realization of the chiral symmetry have been considered which do without the
Higgs mesons. One can think that a linear realization of the chiral symmetry
(at least, for the lightest states and related phenomena) is valid. First,
this is a simple and beautiful mechanism that works also in other fields of
physics, for example, in superconductivity. Second, the effective
Lagrangians obtained on the basis of this mechanism (the Nambu --
Jona-Lasinio and other models) describe perfectly the ground states and
related phenomena.

The analysis of the used experimental data tells that, if the ${f_0}(1370)$
resonance exists (soluiton (2)), it has the dominant $s{\bar s}$ component,
because the ratio of its coupling constant with the $\pi\pi$ channel to
the one with the $K\overline{K}$ channel is 0.12 (as to that assignment of
the ${f_0}(1370)$ resonance, we agree, {\it e.g.,} with the work
\cite{Shakin1}).
A minimum scenario of the simultaneous description of the processes
$\pi\pi\to\pi\pi,K\overline{K}$ does without the ${f_0}(1370)$ resonance.
The $K\overline{K}$ scattering length is very sensitive to the existence of
this state.

The $f_0 (1500)$ state is represented by the pole cluster which corresponds
to a glueball. This type of clusters reflects the flavour-singlet structure
of the glueball wave function and is only a necessary condition of the
glueball nature of the $f_0 (1500)$ state.

We think that multichannel states are most adequately represented by
clusters, {\it i.e.} by the pole positions on all corresponding sheets. The
pole positions are rather stable characteristics for various models, whereas
masses and widths are very model-dependent for wide resonances.

Finally, note that in the model-independent approach, there are many adjusted
parameters (although, {\it e.g.} for the $\pi\pi$ scattering, they all are
positions of poles describing resonances). The number of these parameters can
be diminished by some dynamic assumptions, but this is another approach and
of other value.

This work has been supported by the Grant Program of Plenipotentiary of
Slovak Republic at JINR.  Yu.S. and M.N. were supported in part by the Slovak
Scientific Grant Agency, Grant VEGA No. 2/7175/20; and D.K., by Grant VEGA
No. 2/5085/99.


\end{document}